# Optical brain imaging using a semi-transparent organic light-emitting diode

Bruno F. E. Matarèse [1,2,3]•, Sébastien Roux[4]•, Frédéric Chavane[4]*, John C. deMello[1,5]**

[1]Department of Chemistry, Imperial College London, South Kensington Campus, London, United Kingdom

[2]Department of Haematology, University of Cambridge, Cambridge, United Kingdom

[3]Cavendish Laboratory, University of Cambridge, Cambridge, United Kingdom

[4]Institut de Neurosciences de la Timone (INT), *Centre National de la Recherche Scientifique (CNRS) and* Aix-Marseille Université, Marseille, France

[5]Department of Chemistry, Norwegian University of Science and Technology, Trondheim, Norway

*frederic.chavane@univ-amu.fr

**john.demello@ntnu.no

•these authors contributed equally to the work

**Abstract**

We report optical brain imaging using a semi-transparent organic light-emitting diode (OLED) based on the orange light-emitting polymer (LEP) Livilux PDO-124. The OLED serves as a compact, extended light source which is capable of uniformly illuminating the cortical surface when placed across a burr hole in the skull. Since all layers of the OLED are substantially transparent to photons with energies below the optical gap of the LEP, light emitted or reflected by the cortical surface may be efficiently transmitted through the OLED and into the objective lens of a low magnification microscope ("macroscope"). The OLED may be placed close to the cortical surface, providing efficient coupling of incident light into the brain cavity; furthermore, the macroscope may be placed close to the upper surface of the OLED, enabling efficient collection of reflected/emitted light from the cortical surface. Hence the use of a semi-transparent OLED simplifies the optical setup, while at the same time maintaining high sensitivity. The OLED is applied here to one of the most demanding forms of optical brain imaging, namely extrinsic optical imaging involving a voltage sensitive dye (VSD). Specifically, we carry out functional imaging of the primary visual cortex (V1) of a rat, using the voltage sensitive dye RH-1691 as a reporter. Imaging through the OLED light-source, we are able to resolve small (~ 0.1 %) changes in the fluorescence intensity of the dye due to changes in the neuronal membrane potential following a visual stimulus. Results are obtained on a single trial basis – i.e. without averaging over multiple measurements – with a time-resolution of ten milliseconds.

**Introduction**

Optical brain imaging is a versatile tool for studying brain function at the mesoscale level.[1,2] In a typical measurement, optical access to the brain of an animal is achieved by retracting the scalp and then thinning or removing a portion of the skull and/or the meninges. The animal is then subjected to a controlled sensory stimulus, and changes in brain function are monitored via changes in absorption, scatter or fluorescence close to the surface of the brain.[1] Absorption- and scatter-based measurements monitor 'intrinsic' optical properties of the untreated brain, while fluorescence-based measurements involve 'extrinsic' application of dye molecules [3,4] or genetically encoded sensors[5] to the cortex. A wide range of physiological processes may be probed via optical brain imaging, depending on the illumination wavelength and the choice of indicator (if any).

Haemoglobin in blood is the most significant intrinsic absorber in the brain, and measurements at visible and near infra-red (NIR) wavelengths can reveal changes in blood flow, blood volume and oxygenation[6–8] For instance, changes in blood oxygenation due to neuronal activity may be monitored by illuminating at ~605 nm where oxyhaemoglobin ($HbO_2$) has a substantially weaker absorbance than deoxyhaemoglobin (Hb), while changes in blood volume may be monitored at ~560 nm, an isosbestic point where Hb and $HbO_2$ have equal absorbance (and the measured reflectance signal therefore depends on the total haemoglobin levels).[1] Illumination with polarized light can be used to probe light-scattering due to cell-swelling and other processes unrelated to blood.[9,10]

In intrinsic monitoring, activation of a neuronal population triggers a local consumption of $O_2$, leading to an increase in deoxyhaemoglobin, followed by an increase in blood flow that replenishes the oxyhaemoglobin.[1,11] The resulting changes in absorption typically occur on a second time-scale, so intrinsic brain imaging provides relatively poor temporal information about neuronal activity.[1] Extrinsic imaging by contrast allows brain activity to be monitored on a millisecond time-scale (1-10 ms typically). The most widely used extrinsic indicators are voltage-sensitive dyes (VSDs)[3,4] and calcium indicators,[12] which emit with intensities that depend respectively on the membrane potential of the cells to which they are bound and the concentration of calcium in the neurons during spiking. In practice, extrinsic changes are frequently masked by strong "pulsation artefacts" that arise from changes in blood flow or changes in Hb and $HbO_2$ levels during the cardiac and/or respiratory cycles.[13] Hence, "blue" indicators [14] that absorb at 630 nm, where haemoglobin absorption is relatively weak, are typically used to (partially) suppress the artefacts.

In this paper, we focus on brain-imaging using VSDs, which undergo near instantaneous changes in fluorescence intensity in response to changes in the neuronal membrane potential. We have chosen voltage sensitive dye imaging (VSDI) as a test-application for transparent OLEDs because it is the most technically challenging optical brain imaging modality, requiring the detection of optical signals that are both rapid (ms time-scale) and small (~0.1 %).[14] It may be confidently assumed that, if transparent OLEDs can be successfully used as light-sources for VSDI, then they may also be used for other less demanding brain-imaging techniques.

The VSD dye molecules are applied to the cerebral cortex – the outermost layer of neural tissue in the brain – where they bind to the outer membranes of the neurons, allowing neuronal responses to external stimuli to be detected optically. Since a large fraction of the neuronal membrane area is dendritic, VSD signals reveal information about the synaptic behaviour of a population of neurons. Importantly, they provide information both above and below the spiking threshold,[4,15] whereas calcium indicators reflect only the spiking activity of neurons.[16] Unfortunately, VSDs are prone to photobleaching under illumination by the light-source. And, as noted above, the weak changes in VSD fluorescence are often masked by substantial changes in haemoglobin absorption over the course of the cardiac and respiratory cycles. To suppress such artefacts, it is necessary to use differential measurement procedures that compare the fluorescence behaviour in the absence and presence of a stimulus, taking care to synchronise the start of each trace to the same point in the cardiac and/or respiratory cycle, and to correct for the effects of photobleaching by normalising each trace to a common starting value.[13,17]

Following Ratzlaff and Grinvald,[18] optical brain imaging is typically performed using a macroscope – a low magnification microscope. The macroscope has a relatively high working distance of several centimetres (so the optics can be located some distance above the entrance to the brain cavity), with an objective lens that has a high numerical aperture for efficient coupling of light out of the brain cavity, see Fig. 1. Illumination is typically carried out using epi-illumination (i.e. via the imaging optics, Fig. 1a) or – in the case of intrinsic measurements – side-illumination (Fig. 1b). The former approach requires an additional beam splitter (a dichroic mirror) inside the imaging optics, adding complexity to the optical set-up; it also causes reflective losses (of both the incident and returning light) and results in unwanted vignetting of the acquired images. The latter approach meanwhile requires substantial vertical clearance between the objective lens of the macroscope and the top of the skull, reducing the efficiency of light-collection by the macroscope. Hence, both approaches have drawbacks in terms of convenience and sensitivity. They also preclude the development of mobile, miniaturised imaging systems that could be attached to a live, moving animal.

Traditionally, a broad-band arc-lamp or incandescent lamp (powered by a stabilized DC power supply) has been used as the light-source, together with associated focusing optics, optical filters, light-guides and shuttering.[11] More recently, however, it has become common to use inorganic light-emitting diodes for illumination due to the narrow wavelength range over which they emit and the cold, stable nature of their emission.[19] In a particularly simple set-up reported by Harrison et al.,[20] a ring of inorganic LEDs was mounted on the objective lens of the macroscope, providing a pseudo-monochromatic light-source with good spatial homogeneity without the need for focusing optics, band-pass filters or mechanical shuttering, greatly simplifying the optical set-up relative to lamp-based set-ups.

Unfortunately – even though LED rings provide significant advantages compared to lamp-based illumination – they still have some drawbacks that limit their application. To achieve uniform illumination, a substantial vertical clearance of 15 mm – 40 mm is typically needed between the LED ring and the skull, which means a large working distance is required between the macroscope and the skull (to prevent the detection optics from blocking the LEDs). The high working distance in turn reduces the amount of reflected or emitted light ("returning light") that is coupled into the macroscope, lowering the detection efficiency. In addition, it is often necessary to horizontally offset the LEDs from the perimeter of the brain cavity to prevent them from obstructing the path of the returning light, resulting in further optical losses. The use of LED rings is further complicated by the frequent need to place other equipment close to the brain cavity, e.g. recording chambers or probes for monitoring electrophysiological behaviour, which can make it difficult to place the LED ring in a location that can provide uniform illumination without shadowing. In the case of large-areas, highly curved cortical structures such as the occipital brain region, achieving uniform illumination using horizontally offset LED rings is almost impossible.

To overcome the issues identified above, we propose here the use of transparent organic light-emitting diodes (OLEDs) for optical brain imaging. OLEDs use organic semiconductors based on thin-film conjugated molecules or polymers as emitters. And, like inorganic LEDs, they act as cold (non-incandescent), stable, energy-efficient solid-state light sources.[21] They have simple sandwich structures (see e.g. Fig. 2a), formed in the simplest case by sequentially depositing onto a glass or plastic substrate a first electrode, one or more layers of organic semiconductor, and a second electrode. The electrodes and active layers are usually no more than a few hundred nanometres in height, so the total device thickness is determined in practice by the supporting substrate and any encapsulating layers used to protect against water and oxygen exposure, which together can be as thin as a few hundred microns. Indium tin oxide (ITO) is a typical choice for the first electrode, behaving as an anode with high transparency in the range 400 - 1200 nm,[22] while metal-oxides, metal nanowires and thin-film metals are common choices for the second electrode (the cathode). If the cathode material is sufficiently thin, then it too will be (semi) transparent, rendering the overall device (partially) transparent to wavelengths of light below the energy gaps of the constituent semiconductors.[23,24] If a transparent OLED were to be used as the light-source for extrinsic optical brain imaging, it could be placed directly over the mouth of the brain cavity (Fig. 1c), without blocking the optical path of the returning light to the macroscope. Indeed, it could even be used in place of the glass slide that is typically used to seal the cavity. Hence, both the light-source and the macroscope could be placed in close proximity to the cortical surface, maximising light coupling into and out of the brain cavity.

In this paper we report proof-of-concept measurements that confirm the feasibility of carrying out high sensitivity, single-trial extrinsic brain imaging, using semi-transparent OLEDs in a through-imaging configuration. The successful use of OLEDs for single-trial VSDI – the most challenging form of brain imaging – indicates they will also be effective light sources for other forms of optical brain imaging.

**Results**

Transparent OLEDs were fabricated by sequentially depositing onto ITO-coated glass: 30 nm of the conducting polymer poly(3,4-ethylendioxythiophene : poly(styrenesulfonate) (PEDOT:PSS), 150 nm of the orange-emitting semi-conducting polymer "Polymer Orange" (Merck, PDO-124), 25 nm of calcium and 12 or 18 nm of aluminium (see Methods for fabrication details). To protect the (air-sensitive) constituent materials against water and oxygen ingress, a second glass substrate was sealed over the top of the device using a UV-curable epoxy resin, resulting in the encapsulated sandwich structure shown schematically in Fig. 2a. The light-emitting area – defined by the spatial overlap of the anode and cathode – was 75 mm$^2$ and the total device thickness, including glass substrates and encapsulation, was approximately 2.4 mm. In the chosen architecture, ITO acts as a transparent anode, PEDOT:PSS is an anodic modifier that planarises the ITO and improves hole injection,[25,26] PDO-124 acts as an emitter, calcium is a low work function cathode with good electron-injecting properties, and aluminium is a capping layer that helps prevent oxidation of calcium to non-conductive calcium oxide. When the ITO/PEDOT:PSS anode is biased positively with respect to the Ca/Al cathode, holes and electrons are injected into the bulk of the PDO-124 active layer, where they combine radiatively with the emission of (orange) light.

PDO-124 was chosen for its spectral compatibility with our chosen voltage sensitive dye RH1691 – a red-absorbing dye that has been extensively used for extrinsic brain imaging. Fig. 2b-i shows the solution-phase absorption and emission spectra of RH1691 in ethanol. The dye exhibits substantial absorption and emission from 550 to 650 nm and from 650 to 750 nm, respectively. Hence, by exciting the dye at wavelengths > 600 nm (where Hb and HbO$_2$ exhibit relatively weak absorption, see Fig. 2b-ii), perturbation of the excitation light and dye fluorescence by haemoglobin can be minimised, reducing pulsation artefacts. Fig. 2b-iii shows the electroluminescence emission spectrum of PDO-124 (SO-PPV), which spans the spectral range 540 – 800 nm, allowing it to excite RH1691. Fig. 2b-iv shows the transmittance versus wavelength for two (otherwise identical) OLEDs with 12 and 18 nm Al capping layers, which indicate reasonable transparency across the visible/near-infra-red spectrum. The two devices show relatively flat transmittance spectra above 550 nm, with the thinner Al layer resulting in approximately 50 % higher device transmittance than the thicker layer. Owing to its higher transmittance, the 12-nm-Al device was used for the imaging experiments reported below. (Thinner Al layers could not be used due to poor device stability).

The current-voltage-luminance characteristics of a 12-nm-Al device are shown in Fig. 2c, using 5 kHz square-wave voltage pulses[*] in the range 6 to 30 V (see Methods and Fig. S1). The current density varied super-linearly from 15 mA/cm$^2$ at 6V to 643 mA/cm$^2$ at 30 V, while the optical power density measured through the ITO/PEDOT:PSS electrode varied sub-linearly from 2.5 to 80 μW/mm$^2$, indicating a slight decrease in power efficiency with increasing drive voltage. To place these values in context, power densities at the cerebral cortex of ~20 and ~100 μW/mm$^2$ are typically used for intrinsic and extrinsic imaging, respectively (see Methods). Hence, providing the transparent OLEDs are placed sufficiently close to the cortex, they are sufficiently bright to function as viable light-sources for optical brain imaging. Importantly, the OLEDs remain transparent when they are emitting light as, for instance, can be seen in Figs. 3a,b, which show photographs of an operational transparent OLED placed on top of a sheet of printed text. The two photographs show the device operating at low biases of ~6 V (a) and ~10 V (b), and in both cases the text is clearly visible through the emitting device. (Note, the drive voltage was kept deliberatively low for these photographs to prevent the emitted light from swamping the light reflected by the paper. A much higher drive voltage of ~20 V was used for the measurements described below, and a photograph of the OLED operating under these conditions may be seen in Fig. S2).

---

[*] Driving OLEDs with moderate duty-cycle voltage pulses often results in higher efficiencies and/or operating lifetimes than fixed-voltage operation.

The feasibility of using transparent OLEDs for optical brain imaging was assessed using an anesthetized Brown Norway rat as a target. The bone and dura-matter were surgically removed from a region above the primary visual cortex (V1), exposing a 5 mm x 5 mm region of the cortex. The cortex was stained with the voltage sensitive dye RH1691, rinsed thoroughly with artificial cerebrospinal fluid (aCSF) to wash away unbound dye, and the craniotomy was then closed with transparent agar and a cover glass (see Methods). A semi-transparent OLED was placed 5 mm above the cover glass, with the transparent ITO anode facing towards the brain. With the OLED placed in this orientation, absorptive losses in the Ca/Al electrode affect only the returning light, which for $\lambda > 550$ nm is attenuated by an approximate factor of three relative to a totally transparent device, see Fig. 2b-iv. The OLED was driven using 5 kHz square-wave pulses of amplitude 20 V, resulting in a forward emitted power density of approximately 60 $\mu$W/mm$^2$.

The complete optical set up is shown in Fig. 3c. Fig. 3d shows an image of the cortical surface obtained with the macroscope, using a transparent OLED (operating at 20 V, 5 kHz) placed above the mouth of the brain cavity as the illumination source; also shown in Fig. 3e for reference is the equivalent image obtained by epi-illumination, using a tungsten-halogen lamp that had been band-pass filtered at 540 nm (see Methods). There is a close spatial correspondence between the two images, confirming the suitability of transparent OLEDs for intrinsic brain imaging. The higher contrast in the epi-illuminated image (Fig. 3e) is mostly due to the use of $540 \pm 10$ nm illumination light (as is typical for imaging the vasculature), where both oxy- and deoxy-haemoglobin in the blood vessels absorb strongly relative to the surrounding areas. Absorption by both forms of haemoglobin is substantially weaker in the red region of the spectrum where the majority of the OLED emission occurs, resulting in a lower contrast image (Fig. 3d). For intrinsic imaging of the blood vessels, it would be better to select a green OLED. However, since our primary aim was to extrinsically monitor the activity of neurons underneath the blood vessels, we deliberately selected an orange OLED together with a compatible red-absorbing VSD. To further minimise artefacts due to haemoglobin absorption, the focal plane of the macroscope was set a few hundred microns below the surface vasculature, causing intrinsic features due to the blood vessels to blur out over multiple pixels.[18]

To correct for pulsation artefacts and photobleaching, the following differential measurement procedure was used.[13,27] 12-bit, 10-ms-per-frame monochrome movies of the VSD fluorescence were recorded on a pairwise basis, with and without a visual stimulus. Each high-resolution movie was stored as a $504 \times 504 \times 120$ multiframe image array $\mathbf{M}$, with the first, second and third dimensions corresponding to the row number, column number and frame number, respectively. The array values (denoting the brightness of the pixels) were stored in double precision floating point format to minimise rounding errors during subsequent data processing. The first movie $\mathbf{M}_a$ in each pair was recorded with an 'active' stimulus, while the second movie $\mathbf{M}_b$ was recorded with a 'blank' stimulus, i.e. no stimulus at all. Both movies were initiated with the stimulus switched 'off'. For $\mathbf{M}_a$ the stimulus was switched 'on' at frame 30, while for $\mathbf{M}_b$ it remained off for the full 1200-ms-duration of the movie (see Fig. 4a). Low-noise ('off-state') reference frames were obtained for the two movies by averaging the first $N = 30$ frames, i.e. prior to any stimulation in both conditions, yielding the $504 \times 504$ images $\bar{\mathbf{I}}_a^{1...30}$ and $\bar{\mathbf{I}}_b^{1...30}$. Each movie was then normalised to initial array values of approximately one by dividing every frame by the reference frame – a process known as frame-zero division (FZD).[†] The normalised frames of each movie were then subjected to 2D-Gaussian smoothing to reduce the image noise (at the cost of a drop in spatial resolution). Hence, after data processing, two normalised and spatially smoothed movies, $\widetilde{\mathbf{M}}_a$ and $\widetilde{\mathbf{M}}_b$, were obtained defined by Eqs. (1) and (2), respectively:

$$\widetilde{\mathbf{M}}_a = [\mathbf{M}_a \oslash \bar{\mathbf{I}}_a^{1\ldots30}] * \mathfrak{J} \tag{1}$$

and

$$\widetilde{\mathbf{M}}_b = [\mathbf{M}_b \oslash \bar{\mathbf{I}}_b^{1\ldots30}] * \mathfrak{J}, \tag{2}$$

where $[\mathbf{M} \oslash \mathbf{I}] * \mathfrak{J}$ denotes pixel-by-pixel division of each frame of a 3D multi-frame image array $\mathbf{M}$ by a 2D image array $\mathbf{I}$, followed by convolution with a Gaussian smoothing filter $\mathfrak{J}$. For equivalent active and blank movies that are triggered to begin at the same point in the cardiac or respiratory cycle, the effect of the above procedure is to equalise the physiological pulsation artefacts and bleaching effects in the two movies. Consequently, the differential, normalised response to the visual stimulus

$$\Delta\widetilde{\mathbf{M}} = \widetilde{\mathbf{M}}_a - \widetilde{\mathbf{M}}_b \tag{3}$$

is largely independent of both physiological pulsation artefacts and bleaching effects, allowing the evoked response to be differentiated on both a temporal (frame-by-frame) and spatial (pixel-by-pixel) basis. The FZD normalisation operation removes the trial-by-trial offset variability, while the differential operation removes the effects of photodegradation and physiological artefacts (since phase-locking the trial onset to the heartbeat cycle synchronises heartbeat artefacts in the active and blank movies).

Fig. 5 summarises the data processing procedure graphically, showing the effect of visual stimulation on the single-trial VSD signal. Pseudocolor images of the reference frames $\bar{\mathbf{I}}_a^{1\ldots30}$ and $\bar{\mathbf{I}}_b^{1\ldots30}$ are shown in Figs. 5a and 5b, with a common colour bar for the two images shown on the right. Superimposed on the colour bar are "spread-lines", indicating the fifth percentile (p5), mean ($\mu$) and ninety-fifth percentile (p95) intensity values for the two reference frames. The two reference frames are similar in appearance, but differ slightly in average brightness with $\bar{\mathbf{I}}_a^{1\ldots30}$ and $\bar{\mathbf{I}}_b^{1\ldots30}$ having mean intensities of 654 and 687 counts, respectively. The lower value for $\bar{\mathbf{I}}_a^{1\ldots30}$ (the second movie in the pair, chronologically) is a consequence of photobleaching of the VSD under illumination by the light-source, and confirms the need to normalise the movies before carrying out the differencing operation. The two reference frames showed similar variations in brightness spatially, with the ratio of the maximum intensity to the minimum intensity being just 2.22 in both cases, indicating good uniformity across the imaging area due to the OLED – a key requirement for reliable imaging. No blood vessel features are evident in the images, indicating the surface vasculature has been effectively "blurred-out" by focusing below the surface and by the signal processing described above. (See Fig. S3 for 3D surface plots of the reference frames).

Figs. 5c and 5d show the images $\bar{\mathbf{I}}_a^{50\ldots75}$ and $\bar{\mathbf{I}}_b^{50\ldots75}$ obtained by averaging frames 50 to 75 of $\mathbf{M}_a$ and $\mathbf{M}_b$, respectively. These frames correspond to onset and a part of the plateau of the VSD evoked single-trial response elicited by the visual stimulus as shown in Fig. 4a, and yield averaged images that look similar to the reference frames, except for a slight reduction in mean pixel intensity due to photodegradation (see spread-lines). Carrying out frame zero division (FZD) of $\bar{\mathbf{I}}_a^{50\ldots75}$ and $\bar{\mathbf{I}}_b^{50\ldots75}$ followed by Gaussian filtering yielded the normalised and smoothed images $\tilde{\mathbf{I}}_a^{50\ldots75}$ and $\tilde{\mathbf{I}}_b^{50\ldots75}$ shown in Figs. 5e and 5f. The close resemblance between the two images indicates that neuronal activation had only a weak influence on the fluorescence, and confirms the need for a high dynamic range camera. Nonetheless, the average pixel intensity of the active image was 0.3 % higher than that of the blank image, indicating a statistically significant enhancement in VSD fluorescence in the presence of the stimulus. (The trial-to-trial difference in the mean pixel intensity for equivalent blank measurements was less than 0.1 %, see Fig. S4). Fig. 5g shows the difference image $\Delta\tilde{\mathbf{I}}$ obtained by subtracting $\tilde{\mathbf{I}}_b^{50\ldots75}$ from $\tilde{\mathbf{I}}_a^{50\ldots75}$, while Fig. 5h shows the same data superimposed on an image of the surface vasculature. $\Delta\tilde{\mathbf{I}}^{50\ldots75}$ is positive over the image area, varying from +0.17 % in the least responsive parts of the image area to +0.35 % in the most responsive parts.[‡] Hence, all neurons in the image zone were successfully activated by the visual stimulus, with those on the right side of the image responding most strongly. This is expected given the retinotopic organization of the visual system: two adjacent points of the visual image exciting the retina are projected on two adjacent points in the visual cortex; the whole retina is therefore mapped in an orderly but distorted fashion on the cortex. The position and size of the activation correlate well with the position and size of the displayed visual stimulus according to the retinotopic organisation of the rat V1 and its cortical magnification factor,[28,29] the strongest activity being in the in the temporal part of the left visual cortex where the stimulus was presented (see Fig. 5h) which represents the upper nasal visual field where the stimulus was presented.

---

[†] Using an average of the first 30 frames (instead of the first frame only) has a small effect on the absolute values of the normalised images (resulting in initial values greater than one if photobleaching is occurring) but does not otherwise affect the analysis.

Fig. 4b shows the result of averaging $\widetilde{\mathbf{M}}_a$ and $\widetilde{\mathbf{M}}_b$ over all pixels in each frame, i.e. it shows the mean normalised pixel intensity versus time for the active and blank measurements. The two traces start from initial values of ~1.002 (rather than exactly one) due to the use of $\bar{\mathbf{I}}_a^{1\ldots30}$ and $\bar{\mathbf{I}}_b^{1\ldots30}$ as reference frames for scaling rather than the initial frames of each movie, $\mathbf{I}_a^1$ and $\mathbf{I}_b^1$. The blank trace shows a progressive exponential decay in magnitude over the course of the measurement, attributable to photodegradation of the VSD. The active trace overlaps the blank trace for the first third of the trace, but shows a positive deviation from the blank trace at longer times, due to enhanced fluorescence from the VSD after application of the visual stimulus. The relatively high noise is due to observations being made on unsmoothed single trials, but see the replicability in Fig. S4. From Fig. 4c, which shows the difference in the mean normalised pixel intensity versus time, a latency of approximately ten frames (= 100 ms) can be seen between application of the stimulus and the increase in the VSD fluorescence, consistent with previously measured dynamics of VSD response in the rat.[30] The differential signal saturates at a mean value of around 0.0038 by frame 65, and remains at this broad level for the remainder of the measurement. Hence, with the stimulus present, the VSD fluorescence single-trial signal is approximately 0.35 % higher than without. This range of modulation is in line with typical high quality single-trial VSD signals recorded in the sensory system of anesthetized animals.[30–32]

**Conclusions and outlook**

In conclusion, we have reported successful optical brain imaging using a semi-transparent organic light-emitting diode (OLED). The OLED serves as a compact, extended light source which is capable of uniformly illuminating the cortical surface when placed across a burr hole in the skull. Since all layers of the OLED are substantially transparent to photons with energies below the optical gap of the light-emitting polymer, light emitted or reflected by the cortical surface is efficiently transmitted through the OLED and into the objective lens of a low magnification microscope ("macroscope"). We applied the transparent OLED to fluorescence-based extrinsic imaging using the voltage sensitive dye RH-

---

[‡] Note, to enhance the colour contrast in Fig. 5g and h, we have defined the colormap to run from the 1st percentile to the 99th percentile, i.e. from 0.002 % to 0.003 %, instead of the minimum (x %) to the maximum (y %) as used for the other spatial maps.

1691. The latter was used to monitor the response of the rat V1 to a visual stimulus, with changes in fluorescence (caused by changes in the neuronal membrane potential) as small as 0.1 % being readily detected on a 10-ms timescale. VSD imaging was chosen because of its demanding nature in terms of illumination stability, power and homogeneity, as well as the small amplitude of its evoked response. The successful use of OLEDs for single-trial VSDI – the most challenging form of optical imaging – indicates they will also be effective light sources for other forms of optical brain imaging such as absorbance imaging of intrinsic signals.

A drawback of using OLEDs as light sources for optical brain imaging is their broadband emission spectra, which typically span several hundred nanometres, precluding their use in applications that require narrow-band excitation. Optical filtering could be used to sharpen the emission (taking care to ensure the filter is also able to pass the returning light) but this would result in a substantially diminished light output. On the other hand, OLEDs have a number of other properties that make them very attractive light-sources for optical brain imaging. The thin-film nature of the OLEDs means the device thickness is dominated by the substrate and any encapsulating layers used to protect against oxygen and water ingress. The (glass-based) devices used here had a total thickness of 2.4 mm, but using plastic substrates and encapsulants would allow the thickness to be reduced to a few hundred microns. Hence, the illumination source could be placed in extreme proximity to the cortical surface, providing efficient coupling of incident light into the brain cavity. In addition, by placing the objective lens of the macroscope close to the upper surface of the OLED, efficient collection of reflected/emitted light from the cortical surface could be achieved.

The ability to fabricate OLEDs and OLED arrays over large areas should be especially beneficial for optical brain imaging in larger animals such as primates, where craniotomies can be as large as 40 mm in diameter and the large area and high cortical curvature can make it impossible to achieve uniform illumination using conventional optics, see Fig. S5 and Appendix S1. The flexible, conformal nature of thin-film plastic OLEDs opens up the possibility of embedding (appropriately curved) OLEDs inside the brain cavity, potentially using a pixelated array of different coloured OLEDs to spatially and colorimetrically control the illumination. In the case of light-based stimulation (optogenetic or infra red stimulation of neural activity[33]), this would make it possible to excite spatial patterns of the cortex conforming to the topographic map organization such as retinotopy or orientation maps in V1. (Note, in separate work we have used the same orange emitter PDO-124 for photo-stimulation of opsin-expressing neurons[34]). Furthermore, by developing near-IR-emitting OLEDs, it would be possible to carry out non-invasive imaging and stimulation through the intact skull.[33,35] Arguably the most exciting possibility that arises from the use of OLEDs as compact, transparent light-sources is the prospect of developing fully miniaturised imaging systems, with light sources and sensors integrated into the skull of a mobile animal, which would allow for live in-vivo monitoring of unrestrained animals.

**Materials & Methods**

**Materials:** The orange-emitting semiconducting polymer "Polymer Orange" (Livilux PDO-124) was obtained from PPF (Merck KGaA) and dissolved in chlorobenzene at a concentration of 6 mg/ml. The solution was prepared the day before device fabrication and stirred at room temperature overnight. An aqueous dispersion of the conducting polymer poly(3,4-ethylendioxythiophene : poly(styrene-sulfonate) (PEDOT:PSS, Clevios AI 4083) was obtained from Heraeus, and used directly with no further processing. 10 Ohm/sq. 25-mm by 25-mm ITO-coated glass substrates were obtained from Xin Yan Technology Ltd. The substrates were cleaned by sonicating in acetone and then in soap and de-ionised water for twenty minutes. They were then sonicated in de-ionised water three times for five minutes, in acetone twice for five minutes, and in isopropanol (IPA) for five minutes, before drying under a stream of air. The substrates were treated in an oxygen plasma immediately before use.

**OLED fabrication.** The ITO-coated glass substrates were spin-coated with a 30 nm layer of PEDOT:PSS at a speed of 3500 revolutions per minute (rpm) for 40 s, and then annealed at 150 °C for 20 min. The PEDOT:PSS was then coated with 150 nm of Livilux PDO-124 at a speed of 1200 rpm for 30 s. 25 nm of Ca was then then deposited on top of the SO-PPV by thermal evaporation through a shadow mask at $10^{-6}$ mbar, followed by 12 or 18 nm of Al as a capping layer. The active area of the devices, defined by the spatial overlap of the anode and cathode, was 75 mm$^2$ (10 mm by 7.5 mm). The completed devices were sealed using UV-curable epoxy (KatioBond 45952, Delo) and a 17 mm x 17 mm glass slide for encapsulation. Electrical contact to the electrodes was made via the still-exposed part of the device.

**Electrical characterisation of OLEDs under pulsed operation.** The OLEDs were driven at 5 kHz, using a variable amplitude square-wave DC bias. The square-wave pulses were generated using the variable DC output from a source measure unit (2450, Keithley) and an N-channel MOSFET (IRLZ34NPBF, Infineon) which operated as a low-side switch under the control of a microcontroller (Uno, Arduino). (The gate of the MOSFET was connected to a digital output pin of an Arduino microcontroller, which generated the necessary 5 kHz driving signal, see Fig. S4). The current-voltage characteristics of the encapsulated devices were measured in air at room temperature, using the SMU to power the MOSFET driver and measure the time-averaged current.

**Optical characterisation of OLEDs.** Transmission spectra in the range 400 to 1200 nm were measured with a Shimadzu UV-3101 NIR-Vis-NUV spectrophotometer. Emission spectra under pulsed operation were measured using a fibre-coupled Ocean Optics 2000+ spectrometer. A calibrated amplified Photodiode (OPT101, Texas Instruments) with a gain of $10^6$ V/A connected to a 100 MHz digital oscilloscope board (NI PCI-5112, National Instruments) was used to measure the optical power intensity of the OLEDs under pulsed operation.

**Animal preparation.** A Brown Norway male rat (1.5 months old) was anaesthetized using an intraperitoneal injection of urethane (1.2 grm/kg). The experimental protocol was approved beforehand by the local Ethical Committee for Animal Research and all procedures complied with the French (approval n°A12/01/13) and European regulations on Animal Research as well as the guidelines from the Society for Neuroscience. To gain optical access to the cortical surface above the primary visual cortex (V1), a complete craniotomy (4 mm x 4 mm) and durectomy were performed. The cortex was stained with voltage sensitive dye (RH-1691, Optical Imaging Inc) suspended at 1mg/ml in artificial cerebrospinal fluid (ACSF) for two hours, and then rinsed thoroughly with ASCF. Finally the aperture was filled with a transparent agar-agar/ACSF mixture, and sealed with a glass slide using dental cement to achieve a clear optical access without pulsatory artefacts or desiccation of the cortex.

**Optical setup.** The optical imaging setup matched the general configuration proposed by Ratzlaff and Grinvald, [18] with two camera objective lenses aligned in a face-to-face geometry, see Fig. 3c. The two lenses were separated by a distance of 150 mm. The lower objective lens (f = 50 mm, F/1.2, Nikkor AI-S, Nikon) was located 40 mm from the cortical surface, while the upper objective lens (f = 85 mm, F/1.4, Nikkor AF, Nikon) was located 40 mm from the front lens of a digital camera (Imager 3001, optical imaging Inc.). The magnification of the macroscope was 1.7 (= 85 / 50). The two objective lenses were set to their maximum aperture settings, while the camera was set to its high sensitivity mode with a (binned) resolution of 504 pixels x 504 pixels. A 665 nm long-pass filter (RG665, Schott) was placed in front of the upper objective lens. For OLED illumination, the semi-transparent OLED was placed across the brain cavity, approximately 5 mm above the glass slide used to seal the cavity. For halogen lamp illumination, a beam-splitting dichroic mirror (Omega optical, DRLP 650) was inserted between the two objective lenses, allowing the halogen lamp (Hal 100, Zeiss) to be placed on one side as in Fig. 2a. Light from the halogen lamp was filtered using a heat filter (KG1, Schott) and a 630 nm band-pass filter (Omega optical, 630 nm DF 10), before passing through a mechanical shutter (Uniblitz electronic) onto the dichroic mirror.

**Functional imaging data acquisition.** A 5 x 5 mm window was imaged using a 504-pixel by 504-pixel camera operating at a frame rate of 100 Hz (Dalsa 1M60P, Dalstar). The camera was controlled using the VDAQ data-acquisition system (Optical Imaging Inc). The OLED was placed 5 mm above the recording chamber. A (Labview-driven) microcontroller (Uno, Arduino) was used to power the OLED with 5 kHz square-wave voltage pulses and to ensure synchronization with the image acquisition system and the computer used to generate the visual stimuli (see below). Each acquisition was phase-locked with the animal heartbeat cycle to allow removal of pulsation artefacts. Each acquisition trial lasted 1200 ms (120 frames), at the end of which the OLED was turned off. The surface vasculature was imaged using the green-filtered (540 nm) halogen lamp.

**Visual stimulation.** Stimuli were displayed monocularly at 60 Hz on a gamma corrected LCD monitor (placed a distance of 21.6 cm from the contralateral animal eye plane) covering 100° (W) x 80° (H) of visual angle using Elphy software (Elphy, Unic, Paris). Two types of trials were presented to the animal. In each trial, a grey screen of 0.5 $Cd/m^2$ average luminance was displayed during the first 300 ms; the visual stimulus was then presented for 600 ms, before reverting to a grey screen of 0.5 $Cd/m^2$ for the remaining 300 ms. The visual stimulus used in the stimulated condition was a large white square of 49 $Cd/m^2$ of 40° side (of visual angle) displayed on a grey background (0.5 $Cd/m^2$) in the upper nasal right visual field of the animal. The non-stimulated condition (blank condition) used as a control to contrast the functional images (see data analysis) consisted of a constant grey screen of 0.5 $Cd/m^2$. The inter-trial interval was 6 s, and each type of trial was repeated at least 20 times.

**Acknowledgements** This work was supported by the EU through project FP7-PEOPLE-212-ITN 316832-OLIMPIA .

**References**

[1] E. M. C. Hillman, *J. Biomed. Opt.* **2007**, *12*, 051402.
[2] E. E. Lieke, R. D. Frostig, A. Arieli, D. Y. Ts'o, R. Hildesheim, A. Grinvald, *Annu. Rev. Physiol.* **1989**, *51*, 543.
[3] A. Grinvald, R. Hildesheim, *Nat Rev Neurosci* **2004**, *5*, 874.
[4] S. Chemla, F. Chavane, *Journal of Physiology-Paris* **2010**, *104*, 40.
[5] M. Carandini, D. Shimaoka, L. F. Rossi, T. K. Sato, A. Benucci, T. Knopfel, *Journal of Neuroscience* **2015**, *35*, 53.
[6] R. D. Frostig, S. A. Masino, M. C. Kwon, C. H. Chen, *Int. J. Imaging Syst. Technol.* **1995**, *6*, 216.
[7] A. Devor, S. Sakadžić, V. J. Srinivasan, M. A. Yaseen, K. Nizar, P. A. Saisan, P. Tian, A. M. Dale, S. A. Vinogradov, M. A. Franceschini, D. A. Boas, *J Cereb Blood Flow Metab* **2012**, *32*, 1259.
[8] I. Vanzetta, *Science* **1999**, *286*, 1555.
[9] L. B. Cohen, B. Hille, R. D. Keynes, *The Journal of Physiology* **1969**, *203*, 489.
[10] R. A. Stepnoski, A. LaPorta, F. Raccuia-Behling, G. E. Blonder, R. E. Slusher, D. Kleinfeld, *Proceedings of the National Academy of Sciences* **1991**, *88*, 9382.
[11] L.-D. Liao, V. Tsytsarev, I. Delgado-Martínez, M.-L. Li, R. Erzurumlu, A. Vipin, J. Orellana, Y.-R. Lin, H.-Y. Lai, Y.-Y. Chen, N. V. Thakor, *BioMed Eng OnLine* **2013**, *12*, 38.
[12] R. Homma, B. J. Baker, L. Jin, O. Garaschuk, A. Konnerth, L. B. Cohen, C. X. Bleau, M. Canepari, M. Djurisic, D. Zecevic, in *Dynamic Brain Imaging* (Ed.: F. Hyder), Humana Press, Totowa, NJ, **2009**, pp. 43–79.
[13] S. Takerkart, P. Katz, F. Garcia, S. Roux, A. Reynaud, F. Chavane, *Front. Neurosci.* **2014**, *8*, DOI 10.3389/fnins.2014.00002.
[14] D. Shoham, D. E. Glaser, A. Arieli, T. Kenet, C. Wijnbergen, Y. Toledo, R. Hildesheim, A. Grinvald, *Neuron* **1999**, *24*, 791.
[15] S. Chemla, F. Chavane, *NeuroImage* **2010**, *53*, 420.
[16] D. Xiao, M. P. Vanni, C. C. Mitelut, A. W. Chan, J. M. LeDue, Y. Xie, A. C. Chen, N. V. Swindale, T. H. Murphy, *eLife* **2017**, *6*, e19976.
[17] S. Chemla, L. Muller, A. Reynaud, S. Takerkart, A. Destexhe, F. Chavane, *Neurophoton* **2017**, *4*, 031215.
[18] E. H. Ratzlaff, A. Grinvald, *Journal of Neuroscience Methods* **1991**, *36*, 127.
[19] M. Nishimura, H. Shirasawa, W.-J. Song, *Neuroscience Research* **2006**, *54*, 230.
[20] T. C. Harrison, A. Sigler, T. H. Murphy, *Journal of Neuroscience Methods* **2009**, *182*, 211.
[21] N. Thejo Kalyani, S. J. Dhoble, *Renewable and Sustainable Energy Reviews* **2012**, *16*, 2696.
[22] Y.-H. Tak, K.-B. Kim, H.-G. Park, K.-H. Lee, J.-R. Lee, *Thin Solid Films* **2002**, *411*, 12.
[23] M. Qian, Q. Liu, W.-S. Li, J.-X. Man, Y.-B. Zhao, Z.-H. Lu, *Organic Electronics* **2018**, *57*, 98.
[24] J. H. Yim, S. Joe, C. Pang, K. M. Lee, H. Jeong, J.-Y. Park, Y. H. Ahn, J. C. de Mello, S. Lee, *ACS Nano* **2014**, *8*, 2857.
[25] G. Huseynova, Y. Hyun Kim, J.-H. Lee, J. Lee, *Journal of Information Display* **2019**, 1.
[26] P. A. Lane, P. J. Brewer, J. Huang, D. D. C. Bradley, J. C. deMello, *Phys. Rev. B* **2006**, *74*, 125320.
[27] A. Reynaud, S. Takerkart, G. S. Masson, F. Chavane, *NeuroImage* **2011**, *54*, 1196.
[28] C. Gias, N. Hewson-Stoate, M. Jones, D. Johnston, J. E. Mayhew, P. J. Coffey, *NeuroImage* **2005**, *24*, 200.
[29] S. Roux, F. Matonti, F. Dupont, L. Hoffart, S. Takerkart, S. Picaud, P. Pham, F. Chavane, *eLife* **2016**, *5*, e12687.
[30] G. Palagina, U. T. Eysel, D. Jancke, *Proceedings of the National Academy of Sciences* **2009**, *106*, 8743.
[31] I. M. Devonshire, T. H. Grandy, E. J. Dommett, S. A. Greenfield, *European Journal of Neuroscience* **2010**, *32*, 786.
[32] I. Ferezou, S. Bolea, C. C. H. Petersen, *Neuron* **2006**, *50*, 617.
[33] M. Chernov, A. W. Roe, *Neurophoton* **2014**, *1*, 011011.
[34] B. F. E. Matarèse, P. L. C. Feyen, J. C. de Mello, F. Benfenati, *Front. Bioeng. Biotechnol.* **2019**, *7*, 278.
[35] G. Hong, S. Diao, J. Chang, A. L. Antaris, C. Chen, B. Zhang, S. Zhao, D. N. Atochin, P. L. Huang, K. I. Andreasson, C. J. Kuo, H. Dai, *Nature Photon* **2014**, *8*, 723.

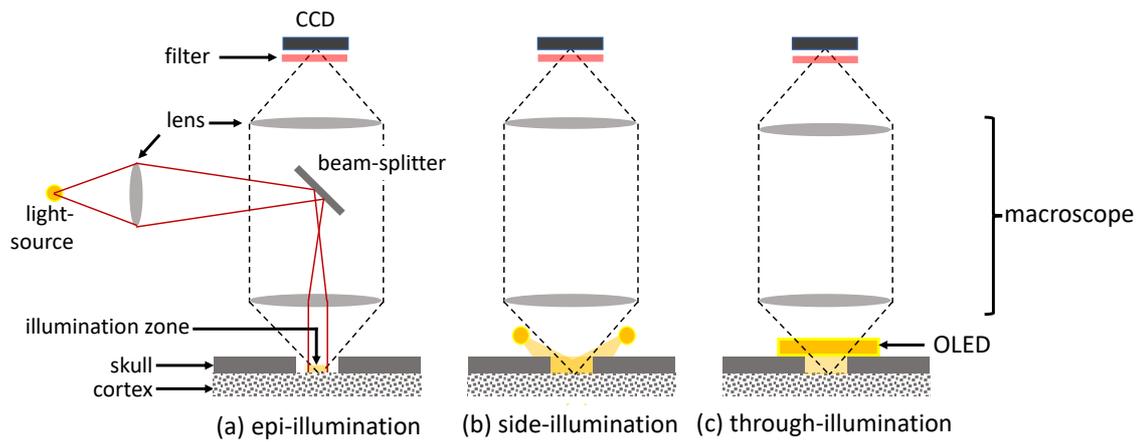

**Fig. 1 – Illumination and detection configurations for optical brain imaging.** In epi-illumination **(a)**, the imaging optics are also used to deliver the excitation light to the cortex; in side-illumination **(b)**, the light-source directly illuminates the cortex, requiring unobstructed optical paths between the light-source and the cortex and between the cortex and the macroscope objective; in through-illumination **(c)**, a semi-transparent OLED located immediately above the brain cavity illuminates the sample and transmits returning light from the sample to the macroscope.

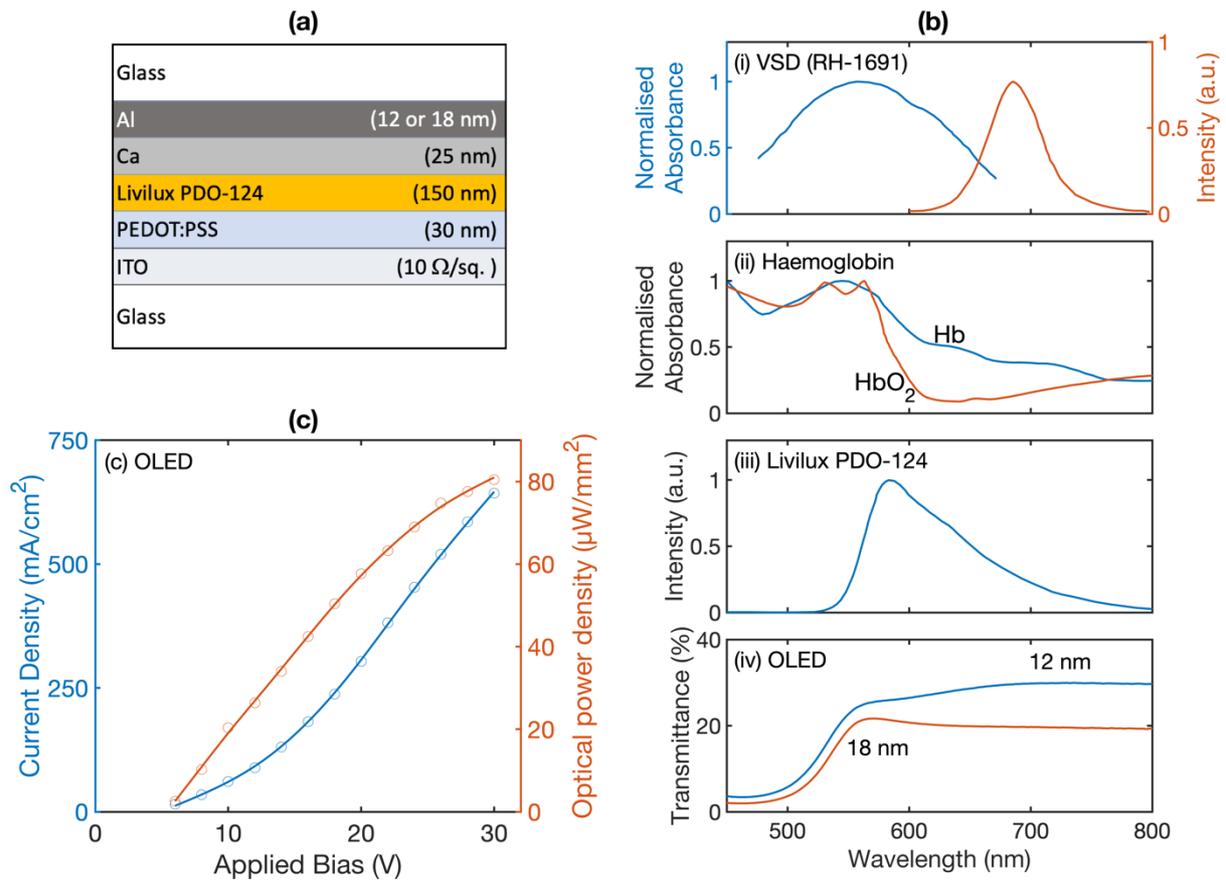

**Fig. 2 – (a)** Structure of the semi-transparent OLEDs. **(b-i)** Absorption (blue) and emission (red) spectra of the voltage sensitive dye RH-1691; **(b-ii)** absorption spectra of haemoglobin [Hb, blue] and oxyhaemoglobin [$HbO_2$, red]; **(b-iii)** emission spectrum of the light-emitting polymer Livilux PDO-124; **(b-iv)** visible light transmittance spectra of semi-transparent OLEDs, fabricated using 12 nm (blue) or 18 nm (red) of Al as a capping layer; **(c)** current-density (blue) and optical power density (red) versus applied bias for semi-transparent OLEDs fabricated using 12 nm of Al as a capping layer.

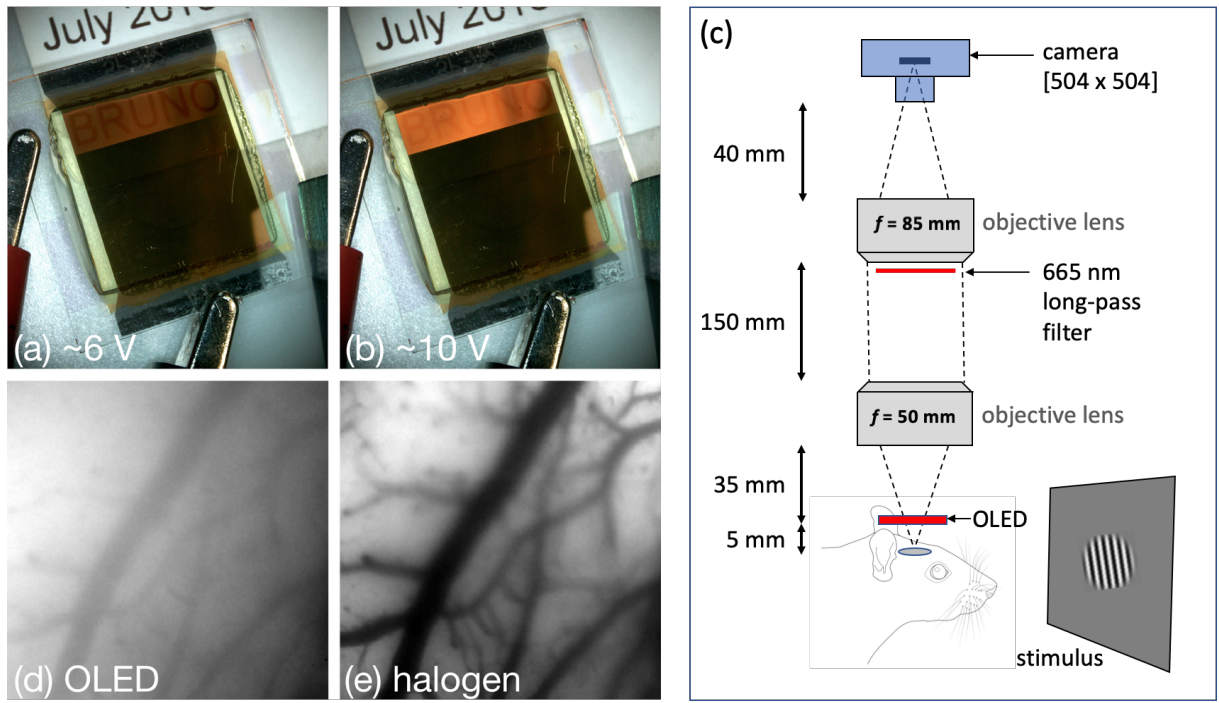

**Fig. 3 – (a, b)** Photographs of a semi-transparent OLED operating under low-voltage/low-brightness conditions; text on a sheet of paper placed beneath the (orange) active area of the device is clearly visible; **(c)** schematic showing the experimental set up used to obtain the intrinsic image in (d) and the extrinsic data in Figs. 4 and 5. (Artwork of rat's head by Gill Brown, CC BY-NC 4.0). **(d, e)** Images of the blood vessels beneath the cortex, obtained using an orange semi-transparent OLED and a green filtered halogen lamp.

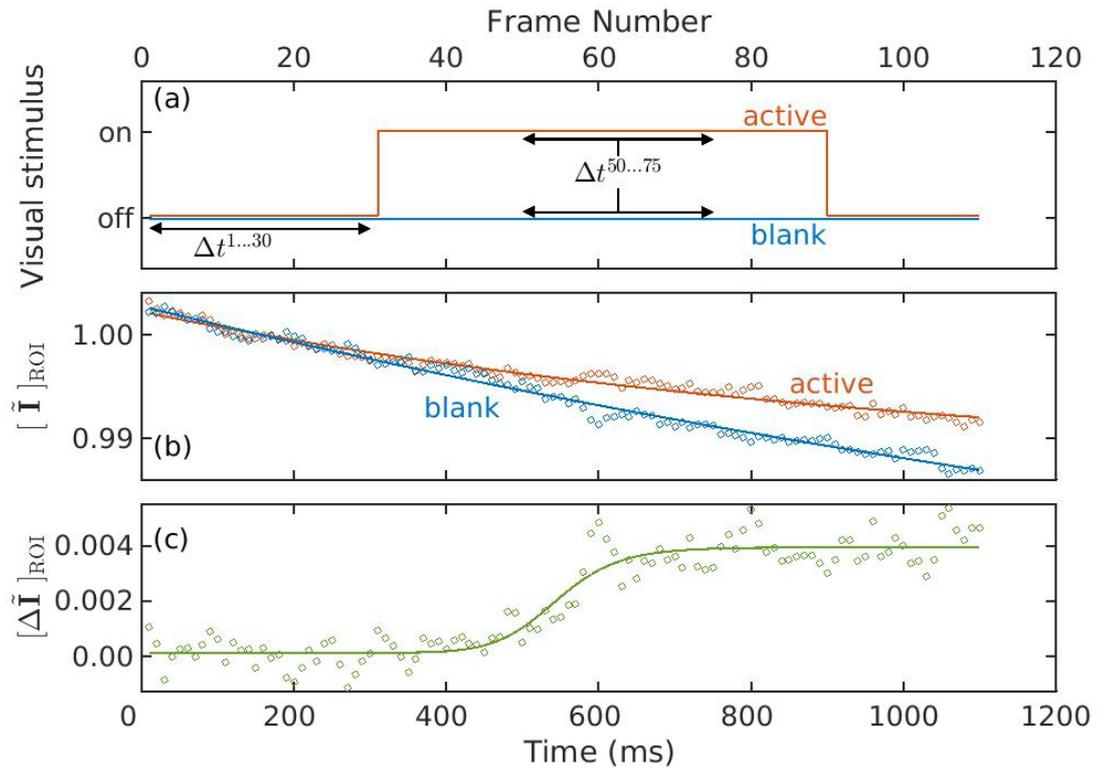

**Fig. 4 – Signal analysis procedure used for extrinsic single-trial measurements on the rat visual cortex, using the voltage sensitive dye RH-1691.** **(a)** Plot showing the on/off state of the visual stimulus versus time for the blank and active measurements; for the active measurement, the visual stimulus is switched off for the first 300 ms, and then switched on up to frame 90; for the blank measurement, the visual stimulus is switched off throughout the measurement; both plots are triggered to start at the same point in the heart beat cycle. **(b)** Mean normalised pixel intensity versus time for the active and blank single-trial measurements, showing a divergence of the signals shortly after the start of the visual stimulus; data is unsmoothed and was acquired on a single-trial basis. Solid lines are guides to the eye. **(c)** Difference in the mean normalised pixel intensity of the single-trial blank and active measurements versus time, obtained using the data in (b). Solid line is a guide to the eye.

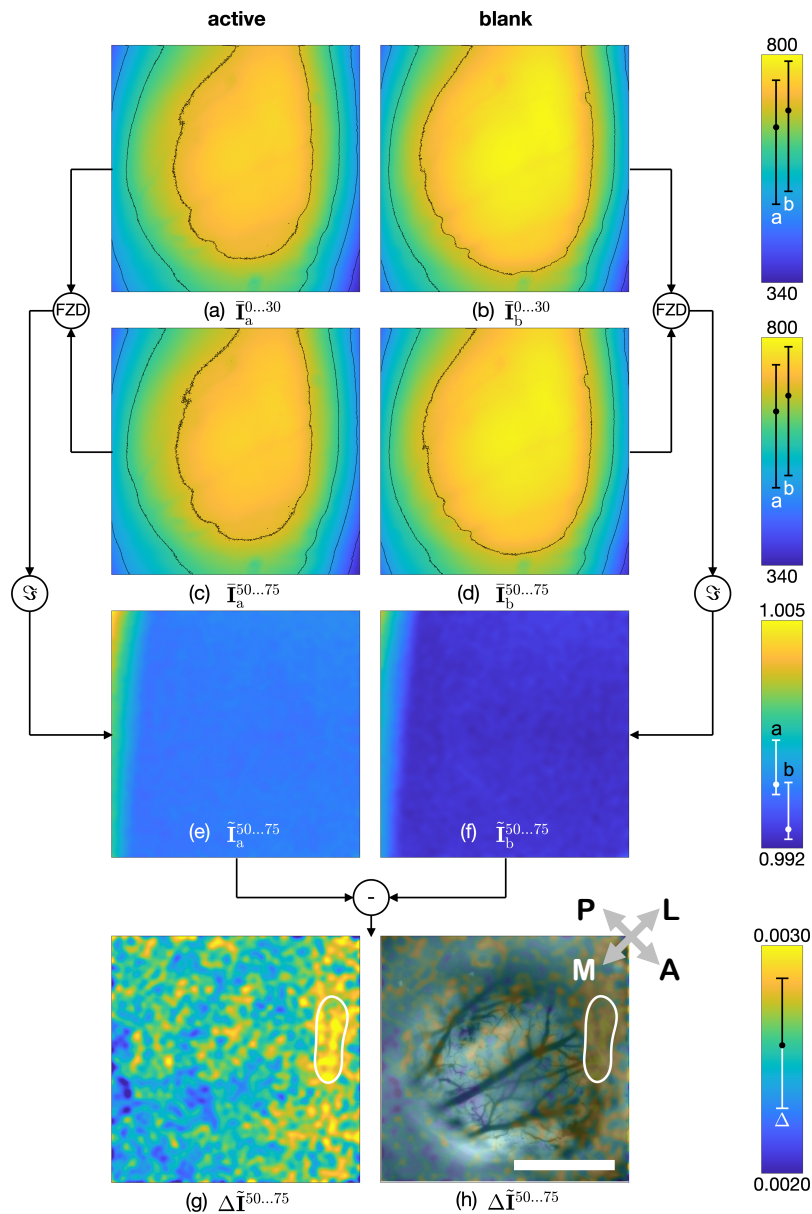

**Fig. 5 – Image analysis procedure used for extrinsic single-trial measurements on the rat visual cortex using the voltage sensitive dye RH-1691. (a, b)** Reference images obtained by averaging the first 30 frames of the blank and active movies during the period where the stimulus is off; a colour bar showing the mapping between colour and pixel intensity is shown on the right hand side; the lines drawn on the colour bar run from the 5th percentile to the 95th percentile, while the solid black circular marker denotes the mean intensity. **(c, d)** Images obtained by averaging frames 50 to 75 of the two movies; in the case of the active measurement, the stimulus is on during this time. **(e, f)** Processed images showing the data from (c, d) after frame zero division (FZD) and Gaussian filtering ($\Im$), see Eq. (1). **(g)** Image obtained by subtracting (f) from (e), showing positive values everywhere due to neuronal activity; the white contour was calculated on a hyper-smoothed version of the functional map and denotes the 85th percentile. **(h)** Image in (g) superimposed on an image of the surface vasculature, showing the strongest activity in the temporal part of the left visual cortex; this region corresponds to the upper nasal right visual field where the stimulus was presented. Scale bar (white): 2 mm, A: anterior, P: posterior, L: lateral, M: Medial.

# Optical brain imaging using a semi-transparent organic light-emitting diode

Bruno F. E. Matarèse [1,2,3]•, Sébastien Roux[4]•, Frédéric Chavane[4]*, John C. deMello[1,5]**

[1]Department of Chemistry, Imperial College London, South Kensington Campus, London, United Kingdom
[2]Department of Haematology, University of Cambridge, Cambridge, United Kingdom
[3]Cavendish Laboratory, University of Cambridge, Cambridge, United Kingdom
[4]Institut de Neurosciences de la Timone (INT), *Centre National de la Recherche Scientifique (CNRS) and Aix-Marseille Université, Marseille, France*
[5]Department of Chemistry, Norwegian University of Science and Technology, Trondheim, Norway
*frederic.chavane@univ-amu.fr
**john.demello@ntnu.no
•these authors contributed equally to the work

**Supplementary Information**

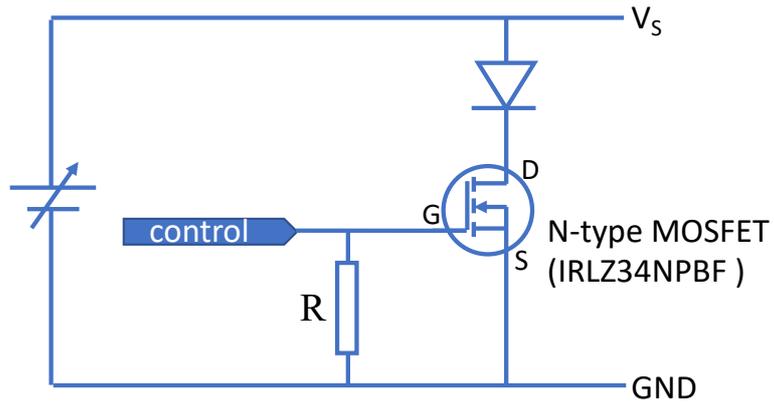

**Fig. S1 Schematic of circuitry used to drive the OLED.** The N-channel MOSFET (IRLZ34NPBF, Infineon) is used as a low-side switch, and its gate is toggled at 5 kHz using the output pin of a microcontroller (Arduino Uno). The voltage across the OLED is controlled using the variable DC output from a source measure unit (2450, Keithley).

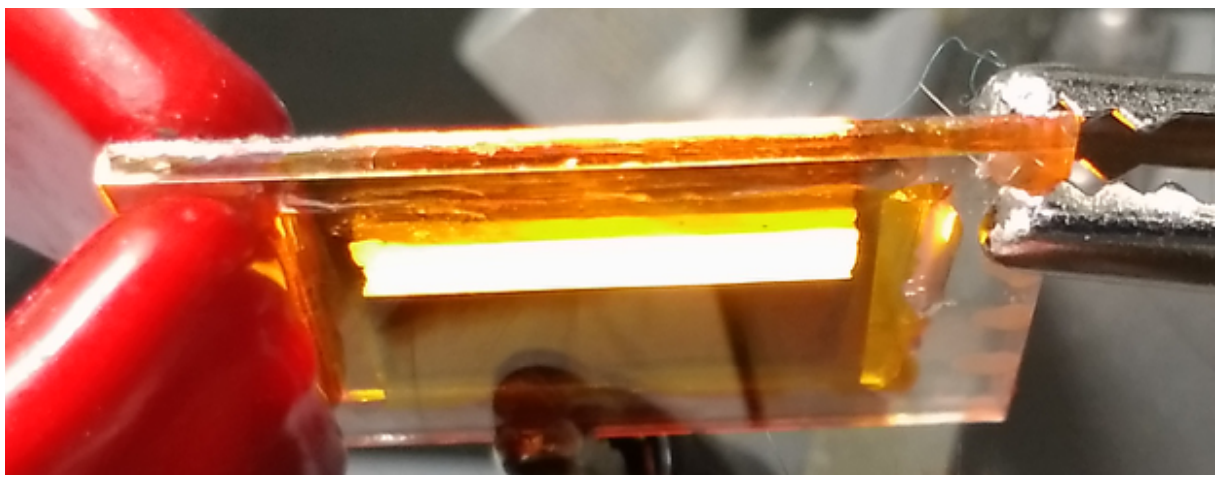

**Fig. S2** Photograph of transparent OLED working under typical operating conditions (20V)

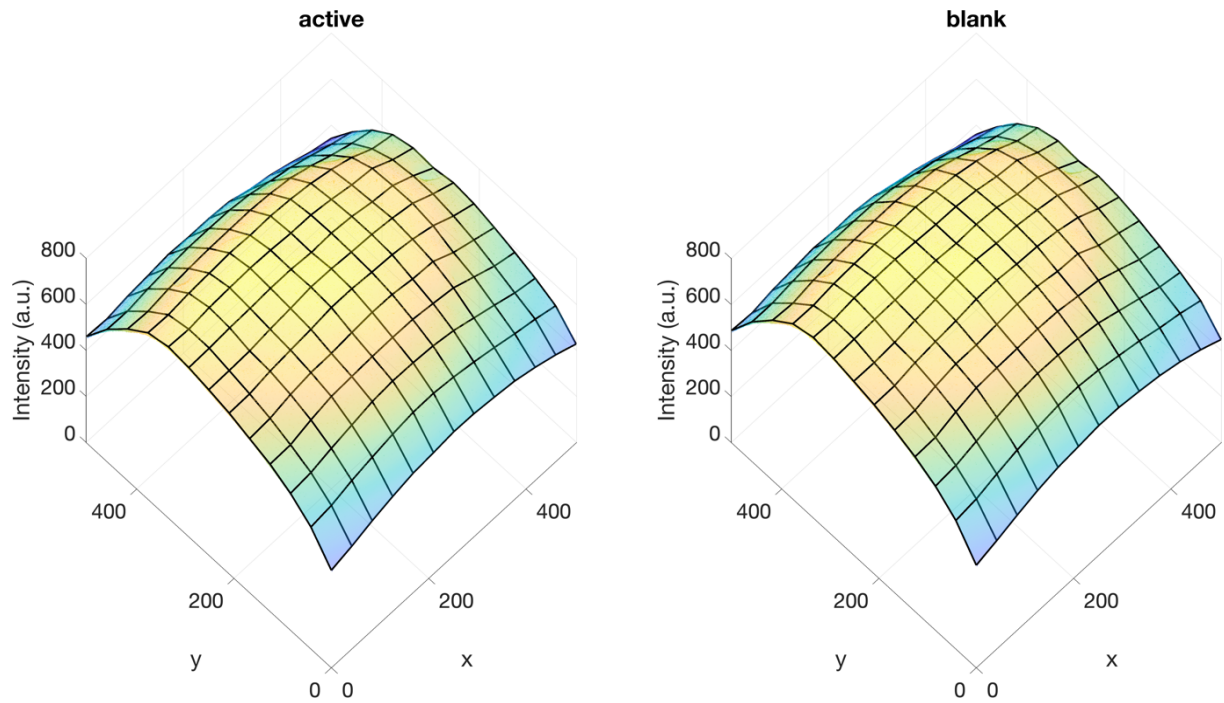

**Fig. S3**  3D surface plots showing the spatial variation in image intensity for reference frames $\bar{\mathrm{I}}_a^{1...30}$ and $\bar{\mathrm{I}}_b^{1...30}$, where x and y represent horizontal and vertical pixel positions. The plots indicate broadly uniform illumination of the surface with no visible dark spots, vignetting or non-uniformities due to the surface vasculature.

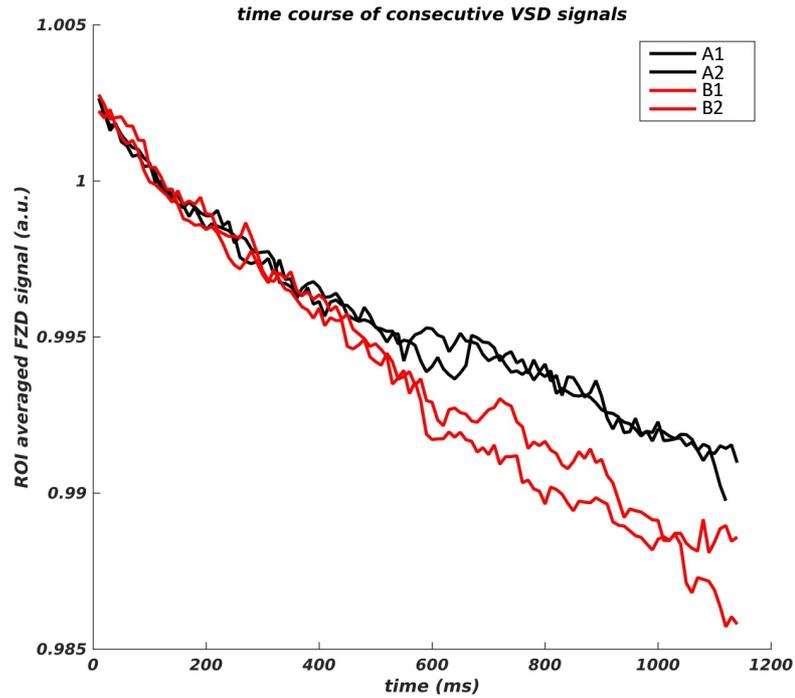

**Fig. S4** Plot showing the mean normalised pixel intensity versus time for sequential active (A1, A2) and blank (B1, B2) single-trial measurements, acquired in the order A1, B1, A2, B2. The average difference between the active and blank signals was 0.3 % between 500 ms and 750 ms (i.e. when the stimulus was active). The difference between the two blank signals was 0.1 % over this same period. The level of jitter in the signals is typical for unsmoothed data acquired on a single-trial basis using a broadband illumination source.

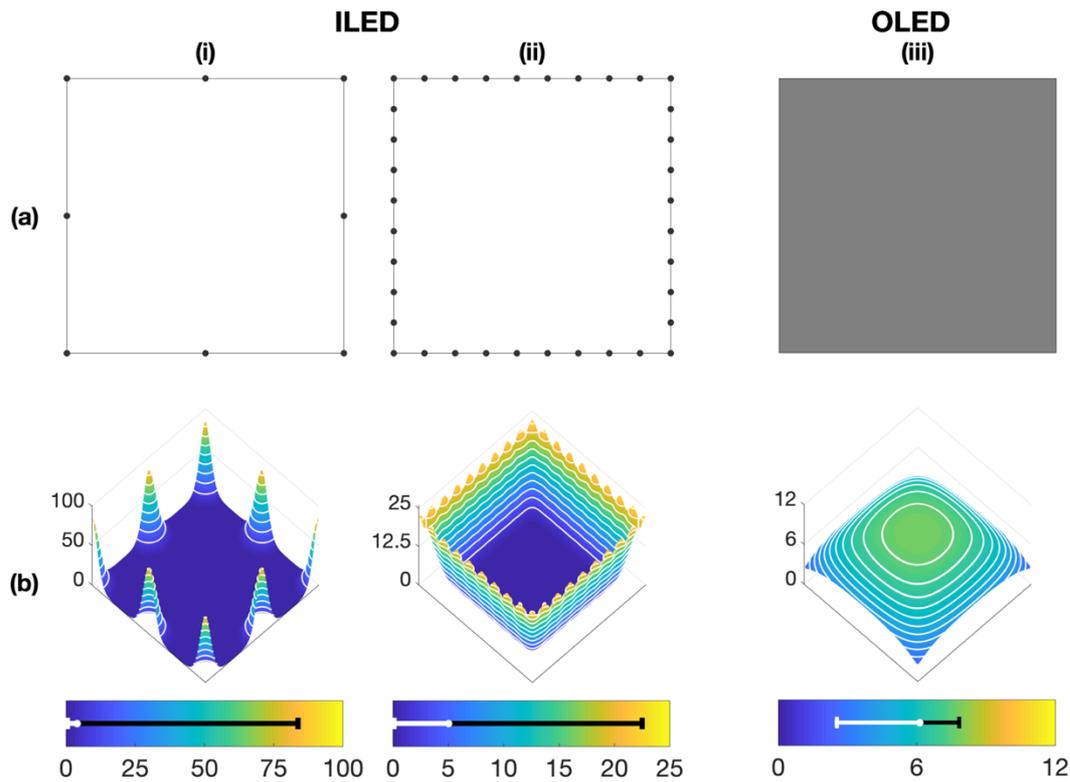

**Fig. S5** **Irradiance patterns on a 15 mm x 15 mm horizontal surface due to a sparse ring of 8 inorganic LEDs (i), a dense ring of 36 inorganic LEDs (ii) and an organic LED of equal area to the surface (iii), each located a vertical distance of 2.5 mm above the surface.** The inorganic LEDs (ILEDs) and the organic LED (OLED) are assumed to have $\theta_{1/2}$ values of 20 degrees and 60 degrees, respectively – the latter value being consistent with OLEDs exhibiting near-Lambertian emission. The total irradiance (obtained by summing the irradiances of the constituent LEDs) is set equal for the three cases. The inorganic LED rings are placed at the perimeter of the surface (**a-i, a-ii**), while the OLED covers the full area of the horizontal surface (**a-iii**). The corresponding irradiance patterns for the different illumination sources are shown in **(b)**, assuming each LED generates an individual irradiance pattern consistent with Eq. (S1) and treating the large area OLED as a dense (100 x 100) square grid of point-like OLEDs. Owing to the close proximity of the ILEDs to the surface and the limited divergence of their 'beams', they illuminate only the edge region of the surface, leaving the interior unilluminated. Since the OLED emits evenly over its active area, the full area of the surface is illuminated. The irradiance generated by the OLED is strongest at the centre of the surface and falls away towards the edges due to an increasing fraction of the light being emitted into the region outside the surface. The colour bars show the range of irradiances generated by each light source, with the horizontal line running from the 5[th] to the 95[th] percentile and the circular marker denoting the mean irradiance. The irradiance patterns illustrate the difficulty of achieving uniform illumination over large areas using ILED rings when the illumination source is placed close to the surface. The uniformity may be improved by increasing the vertical clearance of the ILEDs from the surface but, for the surface to be imaged, a clear optical path must be maintained both between the ILEDs and the surface and between the imaging optics and the surface, which is difficult to achieve in practice. In the case of a semi-transparent OLED, the surface may be straightforwardly imaged through the OLED.

**Appendix S1 - Comparison of the irradiance distributions due to ILED rings and large area OLEDs**

To a good approximation, inorganic LEDs behave as point-like, non-ideal Lambertian emitters with an irradiance distribution $E(r,\theta) = (E_0/r^2)\cos^m\theta$, where $r$ is the distance from the LED, $\theta$ is the viewing angle, $E_0$ is the on-axis irradiance at $r = 1$, and $m \geq 1$ is a constant that determines how quickly the irradiance decreases with viewing angle; $m = 1$ corresponds to perfect Lambertian emission, while higher values indicate a faster drop-off in irradiance with viewing angle, with the irradiance reducing to half its on-axis value at an angle $\theta_{1/2} = \cos^{-1}(\sqrt[m]{0.5})$. For an (incoherent) array of $N$ such emitters, it can be shown that the irradiance $I$ at a position $(x,y)$ on a flat surface in the plane $z = 0$ is given by: [1]

$$I(x,y) = \sum_{i=1}^{N} I_i^0 Z_i^m [(X_i - x)^2 + (Y_i - y)^2 + (Z_i - 0)^2]^{-(m+2)/2} \tag{S1}$$

where $I_i^0$ and $(X_i, Y_i, Z_i)$ denote the intensities and cartesian coordinates of the $i$-th LED. Using this expression, we compare in Fig. S5 the irradiance distributions across a 15 mm x 15 mm surface for three different illumination configurations: a sparse 'ring' of eight LEDs distributed evenly around the edge of the square (Fig. S5ai) ; a dense ring of 36 LEDs (Fig. S5aii); and a 15 mm x 15 mm semi-transparent OLED. The distance between the light-source and the surface is set to 2.5 mm. For the inorganic LEDs we set $\theta_{1/2} = 20°$ – a typical angle of half-irradiance for inorganic LEDs, corresponding to an $m$-value of 11.14; while for the organic LED we set $\theta_{1/2} = 60°$ ($m$ = 1), consistent with the near-Lambertian nature of OLED emission. [2] We treat the large-area OLED as a square array of 100 x 100 point-like OLEDs. All $N$ LEDs in a given configuration have the same brightness $I_0$, and all three configurations have the same total luminance $I_{tot}$, i.e. for each configuration: $\sum_{i=1}^{N} I_i^0 = NI_0 = I_{tot}$.

Since they are in such close proximity to the surface, each inorganic LED is able to illuminate only a small part of the surface, resulting in localised 'pools' of light. Hence, as seen in the surface plot of Fig. S5bi, for a sparse ring of eight LEDs arranged around the edge of the square, only the outer regions of the square are appreciably illuminated, with the irradiance decreasing substantially in the gaps between the LEDs and falling to just 0.0001 % of its maximum value in the centre of the square. The consequence of such uneven illumination would be extreme underexposure of the central part of the image. Increasing the number of LEDs on the ring to 36 (Fig. S4bii) improves the uniformity of light emission around the edge of the square, but has only a weak effect on the irradiance in the centre, increasing it to 0.0003 % of the maximum irradiance. Extreme underexposure would again result in the interior of the square. In the case of the OLED, due to the extended nature of the light source, the full area of the square is illuminated. The irradiance is strongest in the centre of the image and weakest at the corners (where 75 % of the light is emitted into regions outside of the square). The minimum irradiance is approximately 32 % of the maximum irradiance, allowing adequate exposure to be achieved in all parts of the image.

[1]  I. Moreno, M. Avendaño-Alejo, R. I. Tzonchev, Appl. Opt. 2006, 45, 2265.

[2]  C. W. Tang, S. A. VanSlyke, Appl. Phys. Lett. 1987, 51, 913.